\documentclass[a4paper,usenatbib]{mnras}

\usepackage{newtxtext,newtxmath}

\usepackage[T1]{fontenc}
\usepackage{ae,aecompl}

\usepackage{graphicx}	
\usepackage{amsmath}	
\usepackage{amssymb}	
\usepackage{natbib}
\usepackage[nolist]{acronym}

\acrodef{MBH}{massive black hole}
\acrodef{TDE}{tidal disruption event}
\acrodef{GC}{Galactic Centre}
\acrodef{HVS}{hypervelocity stars}
\acrodef{LK}{Lidov-Kozai}

\title[Binaries in galactic nuclei]{Stellar binaries in galactic nuclei: tidally stimulated mergers followed by tidal disruptions}

\author[B. Bradnick et al.]{
B. Bradnick,$^{1}$\thanks{benb@star.sr.bham.ac.uk}
I. Mandel,$^{1}$\thanks{imandel@star.sr.bham.ac.uk}
Y. Levin$^{2}$\thanks{yuri.levin@monash.edu}
\\
$^{1}$School of Physics and Astronomy, University of Birmingham, Birmingham, B15 2TT, United Kingdom\\
$^{2}$Monash Center for Astrophysics and School of Physics and Astronomy, Monash University, Clayton, VIC 3800, Australia\\
}

\date{Accepted XXX. Received YYY; in original form ZZZ}

\pubyear{2016}

\begin{document}
\label{firstpage}
\pagerange{\pageref{firstpage}--\pageref{lastpage}}
\maketitle

\begin{abstract}
We investigate interactions of stellar binaries in galactic nuclear clusters with a \ac{MBH}.   We consider binaries on highly eccentric orbits around the \ac{MBH} that change due to random gravitational interactions with other stars in the nuclear stellar cluster.  The pericenters of the
orbits perform a random walk, and we consider cases where this random walk slowly brings the binary to the Hills tidal separation radius (the so-called empty loss-cone regime). However, we find that in a majority of cases the expected separation does not occur and instead the members of the binary merge together.  This happens because the binary's eccentricity is excited by tidal interactions with the \ac{MBH}, and the relative excursions of the internal eccentricity of the binary far exceed those in its internal semimajor axis.  This frequently reduces the pericenter separation to values below typical stellar diameters, which induces a significant fraction of such binaries to merge ($\gtrsim 75\%$ in our set of numerical experiments). Stellar tides do not appreciably change the total rate of mergers but circularise binaries, leading to a significant fraction of low-eccentricity, low-impact-velocity mergers. Some of the stellar merger products will then be tidally disrupted by the \ac{MBH} within $\sim 10^6$ years.  If the merger strongly enhances the magnetic field of the merger product, this process could explain observations of prompt relativistic jet formation in some tidal disruption events.
\end{abstract}


\acresetall

\section{Introduction}

Dozens of \ac{TDE} candidates have been observed at a variety of wavelengths, including X-rays \citep{ref:Komossa_Greiner_1999,ref:Komossa_et_al_2004,ref:Levan_et_al_2011,ref:Burrows_et_al_2011}, UV \citep{ref:Gezari_et_al_2009,ref:Bloom_et_al_2011}, optical \citep{ref:van_Velzen_et_al_2011,ref:Gezari_et_al_2012,ref:Chornock_et_al_2014,ref:Holoien_et_al_2014,ref:Arcavi_et_al_2014} and radio \citep{ref:Zauderer_et_al_2011}. The physics of tidal disruptions, including the theoretical investigation of \ac{TDE} lightcurves, have been explored by \citet{Rees:1988,ref:Phinney_1989,ref:Magorrian_Tremaine_1999,ref:Lodato_et_al_2009,ref:Strubbe_Quataert_2009,ref:MacLeod_et_al_2012,ref:Guillochon_Ramirez-Ruiz_2013,ref:Shen_Matzner_2014,ref:Shiokawa_et_al_2015,ref:Bonnerot_et_al_2016} and others.

The majority of stars are members of stellar binaries. Binaries in a galactic nuclear cluster can scatter off other objects in the dense ($\gtrsim10^6\:\mathrm{pc^{-3}}$) stellar cluster \citep{ref:Spitzer_Hart_1971,ref:Antonini_et_al_2010} surrounding the central \ac{MBH} onto highly eccentric orbits around the \ac{MBH}.  Tidal interactions with the \ac{MBH} can then separate the binary; in the classical picture, one component may be ejected as a hypervelocity star \citep{ref:Hills_1988,ref:Yu_Tremaine_2003,ref:Gualandris_et_al_2005,ref:Brown_et_al_2005,ref:Sari_et_al_2010,ref:Brown_2015}, while the other may be subsequently tidally disrupted by the \ac{MBH}. 

\citet{ref:Mandel_Levin_2015} investigated binaries which were scattered toward the \ac{MBH} from large radii.  A single scattering event could move such binaries onto nearly radial orbits around the \ac{MBH}, fully populating the loss cone around the \ac{MBH} \citep{ref:Lightman_Shapiro_1977}.  The component stars of such binaries may be tidally disrupted directly following the tidal separation of the binary, leading to double \acp{TDE}; \citet{ref:Mandel_Levin_2015} estimated that 5 to 10 percent of all \acp{TDE} could be double \acp{TDE}.  Only about 6\% of the simulated binaries in the full loss cone were brought to merger by tidal interactions with the \ac{MBH}.  

We complete the earlier work of \citet{ref:Mandel_Levin_2015} by considering binaries that are scattered toward the \ac{MBH} from smaller radii.  The reduced lever arm means that such binaries cannot immediately transition onto nearly radial orbits, i.e., into the loss cone for tidal separation, but may instead gradually and stochastically approach the tidal-separation loss cone through small angular momentum changes over many orbits.  Binaries on orbits that pass within a few tidal separation radii from the \ac{MBH} experience tidal perturbations.  These perturbations can significantly change the angular momentum of the inner binary without significantly modifying its energy.   The eccentricity of the inner binary can be driven close to unity, resulting in a binary merger.  In contrast to the low fraction of tidally stimulated mergers of binaries in the full loss cone, we find that 80\% of the empty loss cone binaries which we simulate result in high-eccentricity mergers, with the remainder becoming tidally separated. 

When stellar tides between the binary components are introduced following an equilibrium tide model \citep{ref:Hut_1981}, the merging fraction drops slightly to 75\%, with the remaining binaries being tidally separated. In the presence of stellar tides, about half of the binaries merge at high eccentricity as before, but the other half merge with low eccentricities ($e\lesssim0.2$) due to efficient tidal circularization of the binary's inner orbit. As a result of stellar tides, these binaries merge at lower semimajor axes. 

Mergers of binaries on less eccentric orbits around the \ac{MBH} as a result of \ac{LK} resonances \citep{ref:Lidov_1962,ref:Kozai_1962} and stellar evolution have been previously considered by \citet{ref:Antonini_et_al_2010,ref:Prodan_et_al_2015,ref:Stephan_et_al_2016}.  In particular, \citet{ref:Antonini_et_al_2010} also investigated stellar binaries around an \ac{MBH} with lower eccentricities than in this study, concluding that \ac{LK} oscillations played a dominant role in producing stellar mergers.  However, as we show here, \ac{LK} resonance is suppressed for binaries on highly eccentric orbits around the \ac{MBH} because scattering relaxation interactions with the surrounding stellar cusp change the angular momentum of the binary's centre of mass around the \ac{MBH} on a timescale that is shorter than the \ac{LK} timescale.  Moreover, merger products from lower-eccentricity orbits are less likely to tidally interact with the \ac{MBH} shortly after merger. 

On the other hand, the merger products arising from binaries on very eccentric orbits around the \ac{MBH} can be tidally disrupted by the \ac{MBH} on timescales of a million years or less.  As well as being more massive and appearing rejuvenated, the  merger products can have their magnetic fields strongly enhanced as a result of the merger \citep{ref:Wickramasinghe_et_al_2014, ref:Zhu_et_al_2015} [however, \citet{GuillochonMcCourt:2017,Bonnerot:2016} criticized the \citet{ref:Zhu_et_al_2015} result, arguing that it did not incorporate a satisfactory magnetic field divergence cleaning scheme].  Prompt jets have been observed in the Swift J164449.3+573451 TDE\citep{ref:Bloom_et_al_2011,ref:Burrows_et_al_2011,ref:Levan_et_al_2011,ref:Zauderer_et_al_2011}. Initial large-scale magnetic fields can aid in prompt jet production \citep{ref:Tchekhovskoy_et_al_2014}, but may not be required \citep{Parfrey:2015}.  Tidal disruptions of the products of recent tidally stimulated mergers with amplified magnetic fields may be potential candidates for prompt jet formation.

The structure of this paper is as follows. In section \ref{sec:methods} we provide details of our binary population model, integration methods, the evolution of the angular momentum of the orbit around the \ac{MBH} and the stellar tides model. We present our results in section \ref{sec:results}. We discuss the implications of the results and the limitations of this work in section \ref{sec:discussion}.

\section{Methods}

\label{sec:methods}

We numerically integrate the trajectories of 1000 individual stellar binaries on highly eccentric orbits around an \ac{MBH}, following an approach similar to \citet{ref:Gould_Quillen_2003}. We model the very eccentric orbit around the \ac{MBH} as a parabolic orbit and numerically integrate only the near-periapsis portion of the orbit, over $\sim 20$ inner orbital periods, where tidal effects from the \ac{MBH} on the inner binary are greatest.   We make use of the \emph{IAS15} integrator in the \emph{REBOUND} software package  \citep{ref:Rein_Liu_2012,ref:Rein_Spiegel_2015} (see acknowledgements for more details).   

In between the integrated periapsis passages, the orbit around the \ac{MBH} evolves due to two-body relaxation.  Each system is evolved until the stellar binary is either  tidally separated or undergoes a merger, defined as the stars approaching to a separation smaller than the sum of their radii.    

We evaluate the impact of stellar tides by integrating the same set of 1000 simulations with a simple prescription for equilibrium tides adapted from \citet{ref:Eggleton_Kiseleva-Eggleton_2001}.   The following subsections contain details of the population model.

\subsection{Binary population}

We use the same approach as \citet{ref:Mandel_Levin_2015} to generate the stellar binary properties. The mass $M_1$ of the primary star is generated from the Kroupa initial mass function \citep{ref:Kroupa_2001}, in the range $M_{1}\in[0.1,100]\:\mathrm{M_{\odot}}$. The mass of the secondary is drawn according to the mass ratio $q=M_{2}/M_{1}$ distribution $p(q)\propto q^{-3/4}$, $q\in[0.2,1.0]$, consistent with observations \citep{ref:Duquennoy_Mayor_1991,ref:Reggiani_Meyer_2011,ref:Sana_et_al_2013}. The radius of each main-sequence star is set using the approximate relationship $R_{*}=(M_{*}/\mathrm{M_{\odot}})^{k}\:\mathrm{R_{\odot}}$ \citep{ref:Kippenhahn_Weigert_1994}, with $k=0.8$ for $M_{*}<\mathrm{M_{\odot}}$ and $k=0.6$ when $M_{*}\geq\mathrm{M_{\odot}}$.

The semimajor axis $a$ of the stellar binary is randomly drawn from $p(a)\propto1/a$ \citep{ref:Opik_1924} with an upper limit of $1\: \mathrm{AU}$ and a lower limit set by the requirement that both stars must initially fit within their Roche lobes at periapsis \citep{ref:Eggleton_1983}.  The binary is further constrained to lie in the range of orbital periods $P_{\mathrm{bin}}\in[0.1\:\mathrm{days},1\:\mathrm{year}]$.  The upper limits in these constraints ensure that the binary is sufficiently compact to avoid disruption through interactions with other stars in the nuclear cluster.

The initial eccentricity distribution is based on \citet{ref:Duquennoy_Mayor_1991}, and is determined from the inner orbital period. Binaries with short periods (under 10 days) are placed on circular orbits and the remaining binaries have their eccentricities $e$ drawn from a Gaussian with mean $\mu=0.3$ and standard deviation $\sigma=0.15$. The orbital plane of the inner (binary) orbit is randomly orientated with respect to the outer orbital plane (around the \ac{MBH}), with the mutual inclination drawn uniformly in $\cos{i}\in[-1,1]$ and the argument of periapsis and longitude of ascending node uniformly drawn from $\omega,\Omega\in[0,2\pi]$.

\subsection{Outer orbit around the \ac{MBH}}

We consider binaries which have been scattered onto high eccentricity orbits around the \ac{MBH} of mass $M_\textrm{MBH}=10^6 M_\odot$.    The angular momentum of the binary's orbit around the \ac{MBH} will wander due to interactions with stars in the Bahcall-Wolf cusp.  This cusp, with number density $n(r)\propto r^{-7/4}$, contains $N\sim 10^6$ stars in the \ac{MBH} sphere of influence extending out to $\sim 1\mathrm{pc}$ \citep{ref:Merritt_2004}.  

\citet{ref:Mandel_Levin_2015} explored binaries in the full loss cone for which the typical change $\langle dh \rangle$ in angular momentum during one orbit around the \ac{MBH} was larger than the minimum angular momentum for tidal disruption by the \ac{MBH} \citep{ref:Lightman_Shapiro_1977}:
\begin{equation}
h_{\mathrm{LC}}\sim\sqrt{GM_{\mathrm{MBH}}a\left(\frac{M_{\mathrm{MBH}}}{M_{\mathrm{bin}}}\right)^{1/3}}\, .
\end{equation}
In this work, we instead focus on binaries which live in the empty loss cone and slowly explore the angular momentum space.  This allows the \ac{MBH} to gradually tidally perturb the inner orbit over many orbital passages near the \ac{MBH}.  The typical fractional evolution of the angular momentum per orbit for our population of binaries is $\langle dh \rangle /h_{\mathrm{LC}}\sim0.1$. 

The timescale for two-body relaxation to change the angular momentum by that of a circular orbit ($h_{\mathrm{circ}}=\sqrt{GMr}$) is given by \citet{ref:Spitzer_Hart_1971}:
\begin{equation}
\tau_{\mathrm{relax}}=\frac{v^{3}}{15.4G^{2}nm_{*}^{2}\log{\Lambda}},
\end{equation}
where $m_{*}\sim0.5\:\mathrm{M_{\odot}}$ is the typical stellar mass, $n$ is the local stellar density, $v\sim\sqrt{GM_{\mathrm{MBH}}/r}$ is the typical stellar velocity at distance $r$ from the \ac{MBH} and $\Lambda\sim0.4N$ is the Coulomb logarithm.  The typical change in the angular momentum during one orbit is given by
\begin{equation}
\langle dh \rangle = h_{\mathrm{circ}} \left[\frac{P_{\mathrm{out}}}{\tau_{\mathrm{relax}}}\right]^{1/2}\, ,
\end{equation}
where $P_{\mathrm{out}}$ is the outer orbital period.
We model this angular momentum evolution as a random walk, applying a single isotropically oriented kick to the outer orbit once per passage around the \ac{MBH}. Each directional component of this 3D kick is drawn from $\Delta h_{i}=\mathcal{N}(0,(\langle dh\rangle/\sqrt{3})^2)$. 

We randomly generate an entire trajectory of up to 1000 kicks prior to commencing the integration of a given binary.  We start our simulation when the initial periapsis of the outer orbit is 5 times the tidal separation radius of the binary, defined as $r_{\mathrm{p,out}}=5R_{\mathrm{TS}}\approx 5a (M_{\mathrm{MBH}}/M_{\mathrm{bin}})^{1/3}$ \citep{ref:Miller_et_al_2005} where $R_{\mathrm{TS}}$ is the tidal separation radius of the binary and $M_{\mathrm{bin}}=M_1+M_2$ is the mass of the binary. The apoapsis distance is drawn to be consistent with the Bahcall-Wolf cusp with $n(r)\propto r^{-7/4}$, within the range $r_{\mathrm{a,out}}\in[100r_{\mathrm{p,out}},1\:\mathrm{pc}]$. This yields a semimajor axis of $a_\mathrm{out}=(r_{\mathrm{p,out}}+r_{\mathrm{a,out}})/2$ and a minimum initial eccentricity of $e_{\mathrm{out}}>0.98$.  A kick trajectory is accepted if it results in the periapsis of the outer orbit reaching the tidal disruption radius of the binary's hypothetical merger product.

\subsection{Stellar tides}
\label{subsec:inner_binary_tides}

Many of our simulated binaries are on close orbits where stellar tides can efficiently circularise the orbit.  We implement a simple equilibrium tide model \citep{ref:Hut_1981} between the inner binary stars, using equations formulated in \citet{ref:Eggleton_Kiseleva-Eggleton_2001}. We consider only the quadrupolar distortion of each star due to its inner binary companion, and ignore additional effects from stellar rotation. The periapsis of the inner binary is held constant while the eccentricity evolution follows
\begin{equation}
\frac{1}{e}\frac{de}{dt}=-V_{1}-V_{2},
\end{equation}
where the component $V_{i}$ for star $i$ in the inner binary is calculated as
\begin{equation}
V_{i}=\frac{9}{\tau_{\mathrm{TF},i}}\left[\frac{1+(15/4)e^2+(15/8)e^4+(5/64)e^6}{(1-e^2)^{13/2}}\right],
\end{equation}
where $\tau_{\mathrm{TF,i}}$ is the tidal friction timescale. This timescale depends on the viscous timescale $\tau_{\mathrm{v,i}}$ and can be calculated using the expression \citep{ref:Eggleton_Kiseleva-Eggleton_2001}
\begin{equation}
\tau_{\mathrm{TF},i}=\frac{\tau_{\mathrm{v},i}}{9}\left(\frac{a}{R_{i}}\right)^8\frac{M_{i}^2}{M_{\mathrm{bin}}M_{j}}(1-Q_{i})^2,
\end{equation}
where $M_{j}$ is the mass of the binary companion. The term $Q$ describes the quadrupole deformability of the star, and we adopt a value consistent with an $n=3$ polytrope star, $Q=0.021$ (calculated from the interpolation formula provided in \citet{ref:Eggleton_et_al_1998}). An estimate of the viscous timescale can be obtained based on the convective turnover timescale \citep{ref:Zahn_1977} and includes a factor from integrating the square of the rate-of-strain tensor of the time-dependent velocity field over the star \citep{ref:Eggleton_Kiseleva-Eggleton_2001}, $\gamma_{i}$:
\begin{equation}
\tau_{\mathrm{v},i}=\frac{1}{\gamma_{i}}\left(\frac{3M_{i}R_{i}^{2}}{L_{i}}\right)^{1/3}.
\end{equation}
We set the value $\gamma_{i}=0.01$ as in \citet{ref:Eggleton_Kiseleva-Eggleton_2001}, and the luminosity of the star, $L$, is determined from the mass-luminosity relationship $L/\mathrm{L_{\odot}}=\alpha(M/\mathrm{M_{\odot}})^\beta$, with the parameters $\alpha$ and $\beta$ provided in Table \ref{table:mass_lum_rel}.

\begin{table}
\begin{tabular}{lll}
\hline
Mass ($\mathrm{M_{\odot}}$) 	& $\alpha$		& $\beta$	\\
\hline
$<0.43$					& $0.23$		& $2.3$	\\
$0.43-2.00$				& $1.00$		& $4.0$	\\
$2.00-20.00$				& $1.50$		& $3.5$	\\
$20.00<$					& $2700.00$	& $1.0$	\\
\hline
\end{tabular}
\caption{Parameters used in the mass-luminosity relationship to obtain the approximate luminosity for a range of stellar masses (\citet{ref:Duric_2012}, \citet{ref:Salaris_Cassisi_2005}).}
\label{table:mass_lum_rel}
\end{table}

As listed in Table \ref{table:stellar_timescales}, the tidal evolution timescale is much greater than the periapsis passage timescale for our binaries.  Therefore, we only include tidal evolution by adjusting the binary parameters between periapsis passages around the \ac{MBH}.  We calculate the median tidal evolution timescale $\tau_{e}=e/(de/dt)$ at two different fiducial values of the eccentricity to show the strong dependence on eccentricity.  This timescale is longer than the $\sim 10^4$-year outer orbital period for the majority of our binaries until the inner eccentricity reaches $e\sim0.8$; above this value stellar tides become efficient in circularising half of the population over the course of an outer orbital period.  We perform this tidal evolution using a fourth-order Runge-Kutta Cash-Karp method.

\begin{table}
\begin{tabular}{lll}
\hline
Timescale				&	Length (years)			& Scaling								\\
\hline
$P_{\mathrm{bin}}$		&	$\sim10^{-2}$	& $\propto a^{3/2}M_{\mathrm{bin}}^{-1/2}$				\\
$\tau_{\mathrm{v}}$		&	$\sim10^{2}$	& $\propto M^{(1-\beta)/3}R^{2/3}\alpha^{-1}$ 			\\
$P_{\mathrm{out}}$		&	$\sim10^{4}$	& $\propto a_{\mathrm{out}}^{3/2}M_{\mathrm{MBH}}^{-1/2}$	\\
$\tau_{\mathrm{LK}}$		&	$\sim10^{5}$	& $\propto P_{\mathrm{out}}^{2} P_{\mathrm{bin}}^{-1}(1-e_{\mathrm{out}}^2)^{3/2}$				\\
$\tau_{\mathrm{relax}}$	&	$\sim10^{9}$	& $\propto M_{\mathrm{MBH}}^{3/2}a_{\mathrm{out}}^{-3/2}n^{-1}m_{*}^{2}\log{0.4N}$ 							\\
$\tau_{e=0.5}$			&	$\sim10^{7}$	& See section \ref{subsec:inner_binary_tides}			\\
$\tau_{e=0.9}$			&	$\sim10^{3}$	& "											\\
\hline
\end{tabular}
\caption{A comparison of relevant timescales relating to the binary. $P_{\mathrm{bin}}$ is the binary orbital period. The eccentricity evolution timescale is the median value for a given eccentricity.}
\label{table:stellar_timescales}
\end{table}

\section{Results of numerical integrations}
\label{sec:results}

\begin{figure}
\centering
\includegraphics[width=\linewidth]{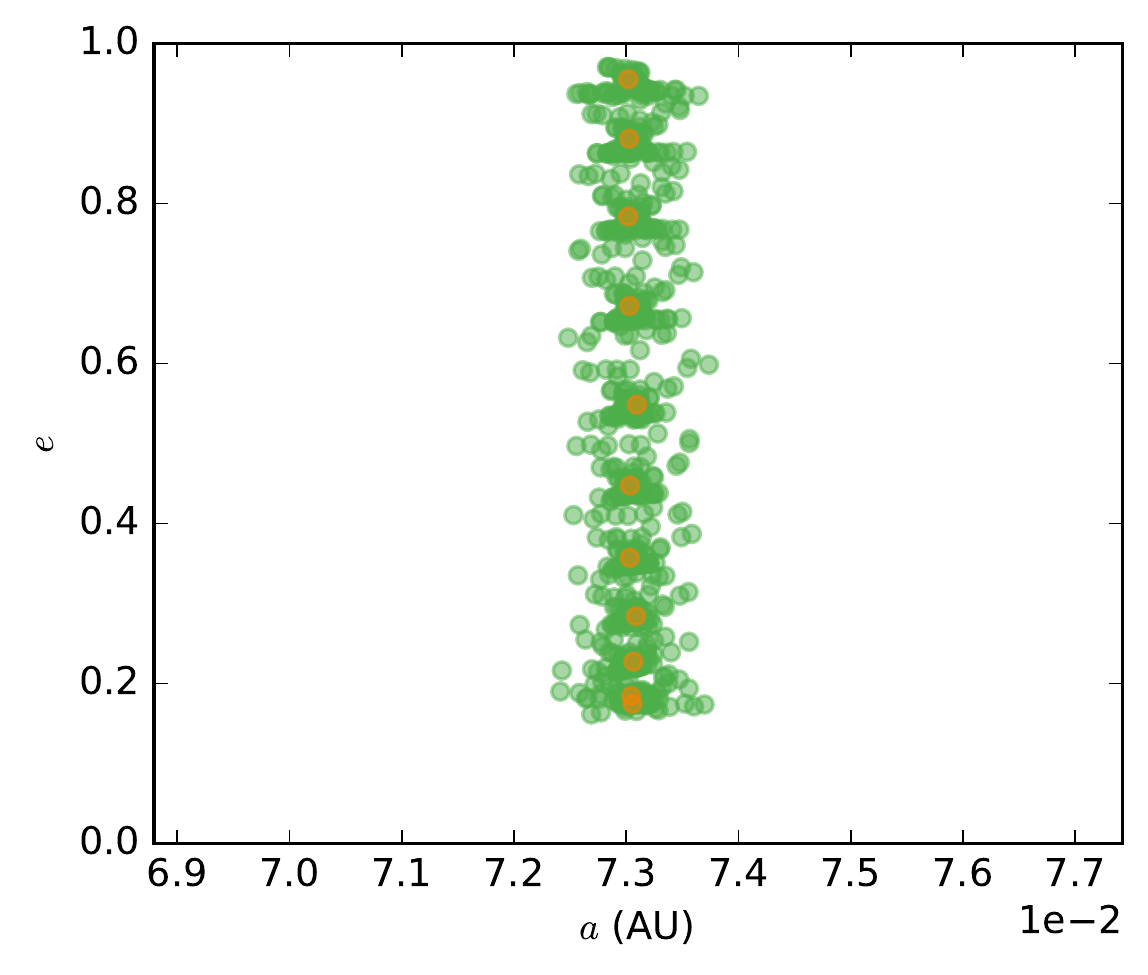}
\caption{Typical evolution of a stellar binary gradually approaching the empty loss cone. The eccentricity is able to grow from an initially small value $e\sim 0.2$ to $e\gtrsim0.95$ whilst the semimajor axis only varies slightly. Orange points mark the initial parameters of the inner binary for each orbit around the MBH, and the green points show the fluctuations during the near-periapsis passages.}
\label{fig:sma_vs_ecc}
\end{figure}

The internal angular momentum of stellar binaries gradually approaching the boundary of the loss cone around the \ac{MBH} typically walks much more than their internal energy.    Gradual tidal interactions between the stellar binary and the \ac{MBH} cause only relatively small fluctuations in the inner binary's energy:
\begin{equation}
\frac{\delta E}{E} \sim \frac{\delta a}{a} \sim \frac{M_\mathrm{BH}}{m_\mathrm{bin}} \frac{a^3}{r_\mathrm{p,out}^3} \sim \left(\frac{R_\mathrm{TS}}{r_\mathrm{p,out}}\right)^3\, .
\end{equation}
On the other hand, the eccentricity of the inner binary is able to evolve efficiently through tidal torques.  Figure \ref{fig:sma_vs_ecc} illustrates the typical evolution of a stellar binary; the eccentricity can be seen to grow from an initially small value of $e\sim 0.2$ to $e \gtrsim 0.95$ while the semimajor axis stays constant to within $\pm 1\%$.  

These eccentricity fluctuations do not represent classical \ac{LK} resonances.  The \ac{LK} timescale for $M_\mathrm{bin} \ll M_\mathrm{MBH}$ and $1-e_\textrm{out} \ll 1$ is \citep{ref:Antognini_2015}:
\begin{equation}
\tau_{\mathrm{LK}}\approx\frac{8}{15\pi}\frac{P_{\mathrm{out}}^{2}}{P_{\mathrm{bin}}}(1-e_{\mathrm{out}}^{2})^{3/2} \approx 0.5 \frac{r_\mathrm{p,out}^{3/2}}{r_\mathrm{TS} ^{3/2}}  \frac{P_{\mathrm{out}}^{2}}{P_{\mathrm{bin}}}.
\end{equation}
Substituting in the initial periapsis of the systems we are investigating, $r_\mathrm{p,out} = 5 r_\mathrm{TS}$, yields a \ac{LK} timescale which is $\sim5$ times the orbital period around the \ac{MBH} (cf.~Table \ref{table:stellar_timescales}).  Therefore, the angular momentum kicks that the binary's orbit around the \ac{MBH} receives at apoapsis due to 2-body relaxation destroy the coherence required for \ac{LK} resonance and suppress classical \ac{LK} oscillations.  

\begin{figure}
\centering
\includegraphics[width=\linewidth]{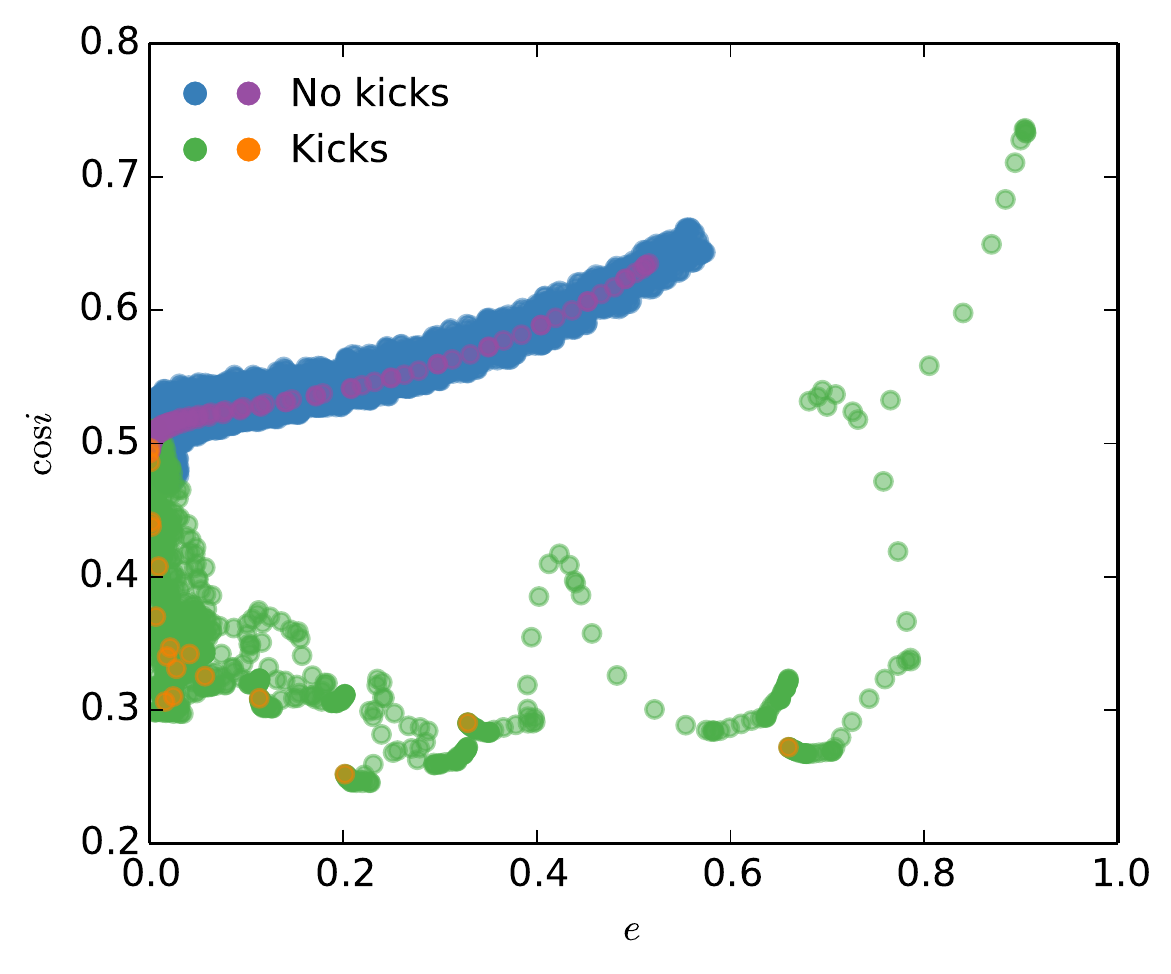}
\caption{The evolution of the inner orbit of a typical stellar binary, showing the eccentricity and cosine of inclination over multiple trajectories around the \ac{MBH}. Evolution is shown both with and without the angular momentum kicks to the outer orbit due to two-body relaxation. The primary colours show the elements throughout each trajectory around the \ac{MBH}, whereas the secondary colours show the initial value for each periapsis passage.}
\label{fig:ecc_vs_inc}
\end{figure}

This suppression is demonstrated in Figure \ref{fig:ecc_vs_inc}, where we compare the evolution of two otherwise identical systems, with and without angular momentum kicks applied far from the \ac{MBH}.   
Without these kicks, the evolution is coherent and can be seen to follow \ac{LK}-like behaviour where the projection of the angular momentum of the binary onto the orbital angular momentum axis remains almost constant. When these kicks are included, the evolution of the inner orbital parameters appears chaotic and can reach very high eccentricities in tens of orbits around the \ac{MBH}.

\begin{figure}
\centering
\includegraphics[width=\linewidth]{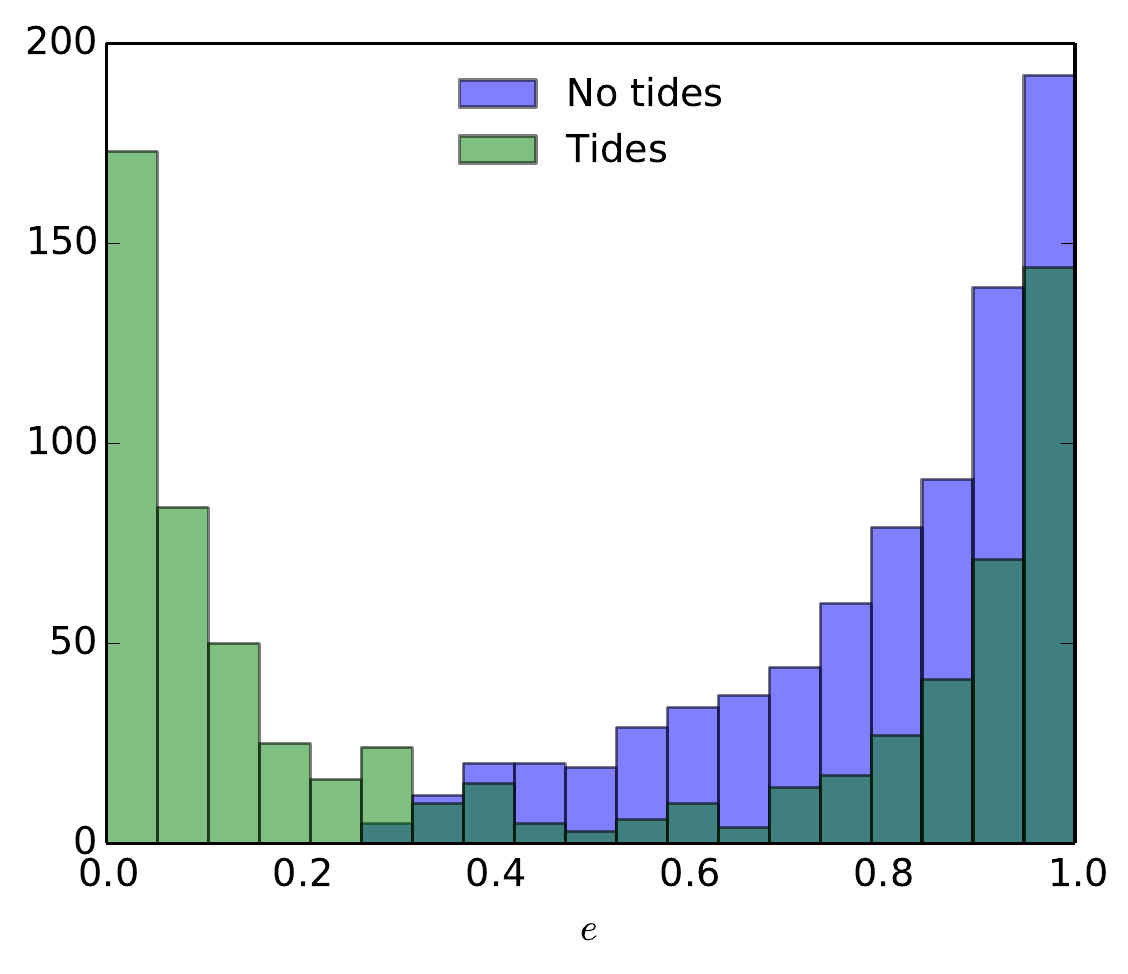}
\caption{Final eccentricity distribution of merging systems, both with and without tides acting between the inner binary components.}
\label{fig:ecc_dist_final}
\end{figure}

\begin{figure}
\centering
\includegraphics[width=\linewidth]{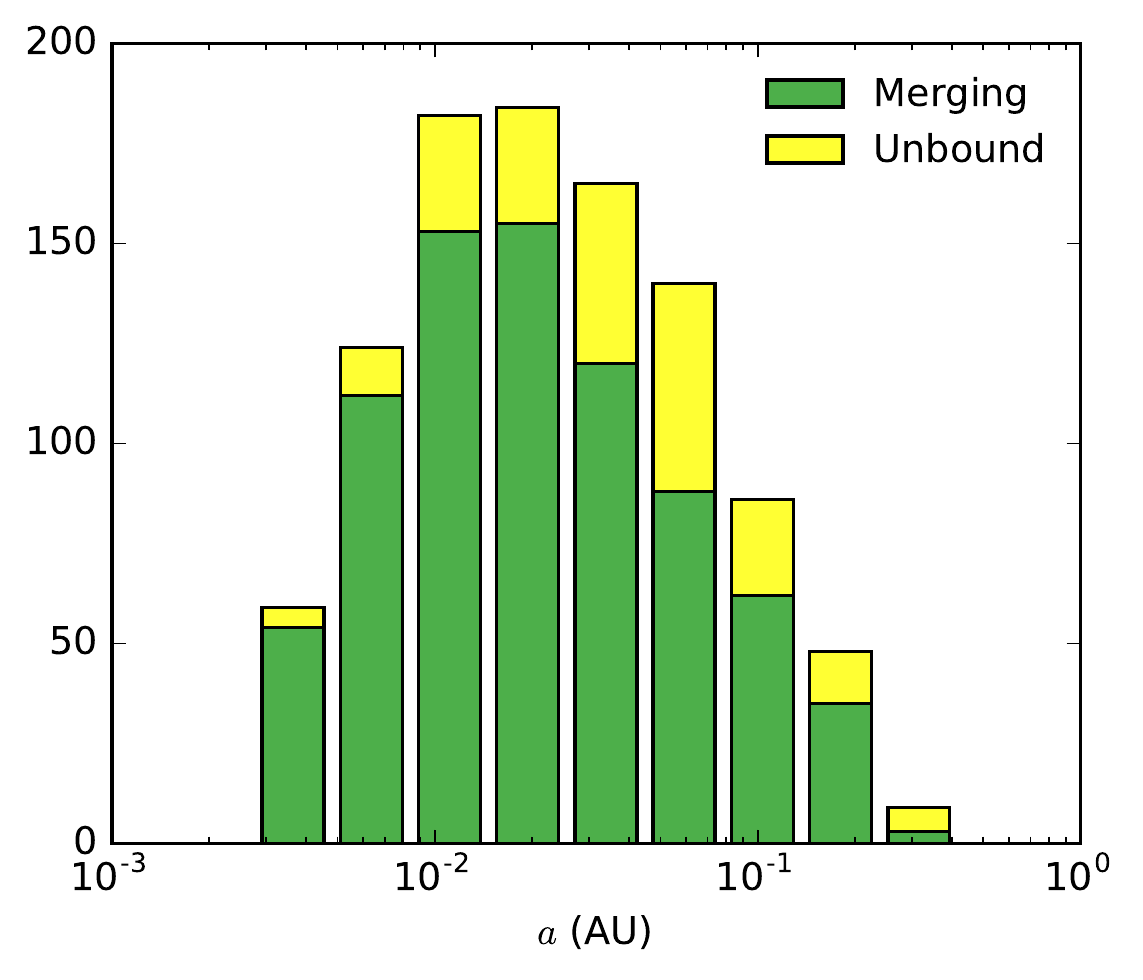}
\caption{Initial binary semimajor axes categorised by the binary's fate (no stellar tides).}
\label{fig:sma_dist_init}
\end{figure}

Such eccentricity excursions lead to a high fraction of tidally stimulated mergers.  The vast majority of the simulated binaries, just under 80\%, result in mergers.  As expected, mergers typically happen at the highest eccentricities, when the inter-star separation at periapsis decreases.  Although $\sim85\%$ of the stellar binaries we simulated are initially circular, the final distribution is ``super-thermal'' (steeper than $p(e) = 2e$), as shown in Figure \ref{fig:ecc_dist_final}. The merger fraction is highest for initially close binaries, which require smaller eccentricity excursions for merger (see Figure \ref{fig:sma_dist_init}).   The remaining binaries are tidally separated, with one of the stars ejected as a hypervelocity star.

The inclusion of stellar tides slightly decreases the overall fraction of mergers to $\sim 75\%$, as high-eccentricity excursions are partially suppressed by tides.   Figure \ref{fig:ecc_dist_final} shows a bimodal distribution of inner binary eccentricities once tides are included, with many binaries circularised by tides.  While tides circularise these binaries, they also harden them; this ensures that merger rather than tidal separation remains the most likely evolutionary outcome.  The dependence of the merger fraction on initial separation with stellar tides remains similar to Figure \ref{fig:sma_dist_init} and is not included here.

\begin{figure}
\centering
\includegraphics[width=\linewidth]{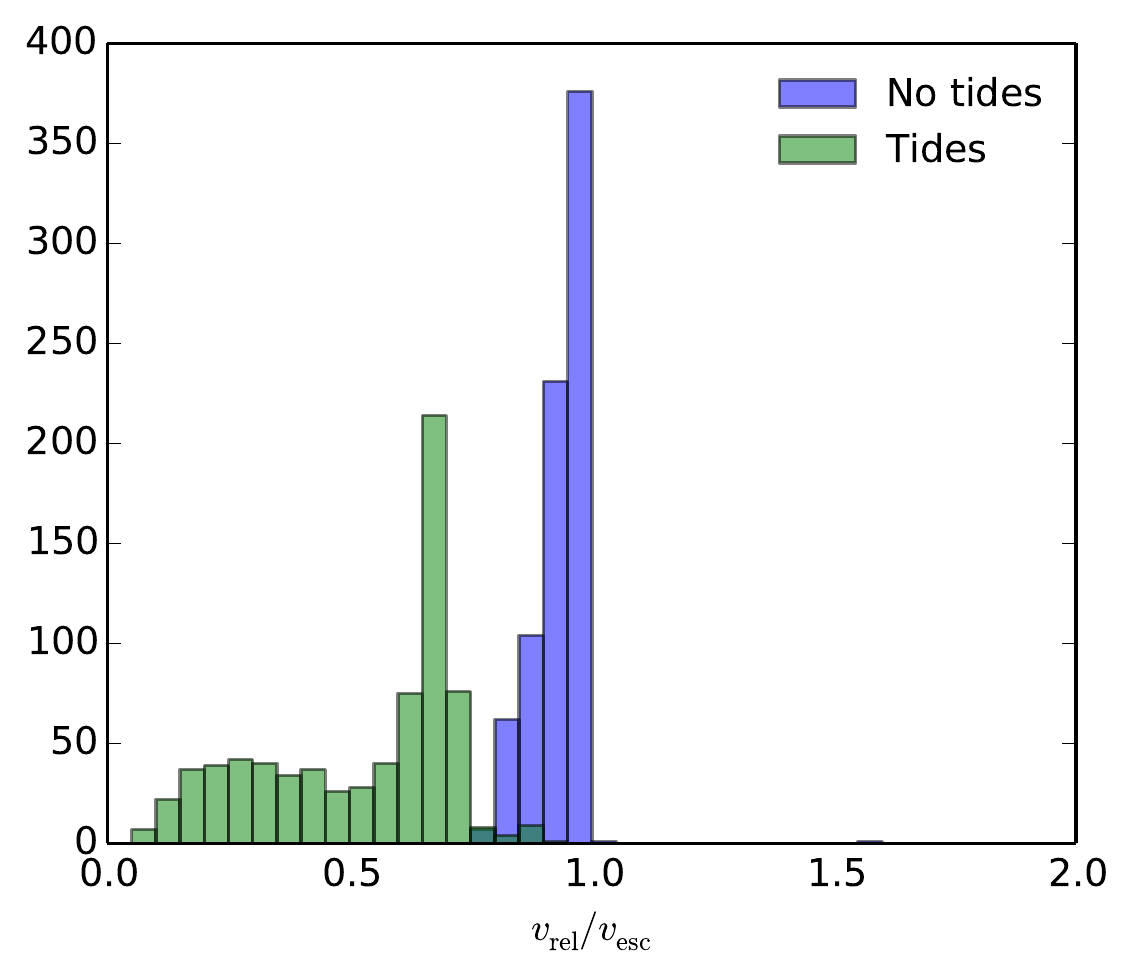}
\caption{Relative velocity at point of impact for merging systems in units of the escape velocity of the primary star, with and without stellar tides.}
\label{fig:v_rel}
\end{figure}

The circularising impact of stellar tides significantly changes the collision dynamics when the merger happens.   The relative velocities between colliding stars are shown in Figure \ref{fig:v_rel}.  Collisions in the absence of stellar tides, which typically happen at a high impact velocity, reaching the escape velocity at the surface of the more massive star, are likely to lead to significant mass loss and an extended merger product.  Meanwhile, grazing collisions in the tidally circularised binaries can have significantly lower relative velocities.

\begin{figure}
\centering
\includegraphics[width=\linewidth]{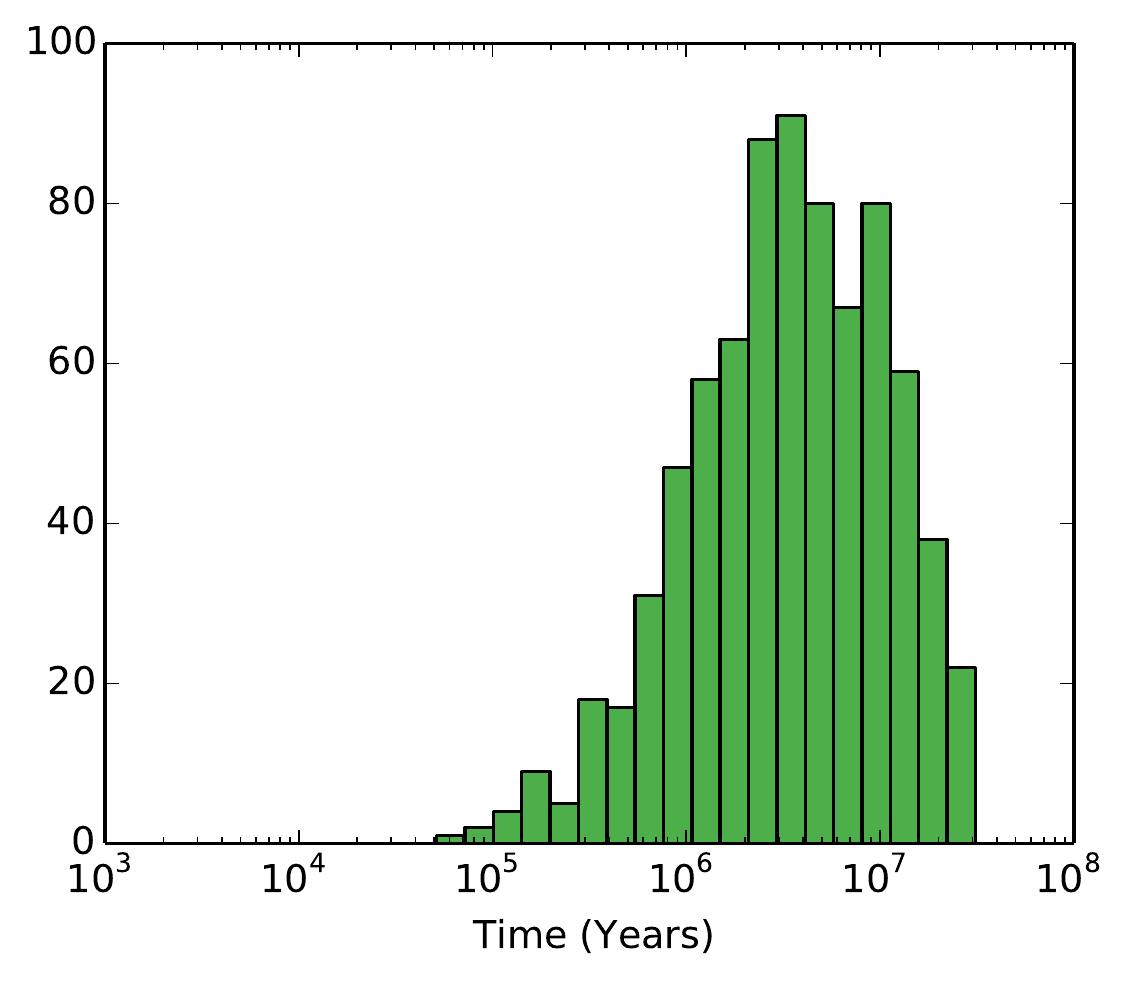}
\caption{Time delay distribution between the tidally stimulated mergers and the tidal disruption of the merger product by the \ac{MBH} in our simulations, chosen so that the merger product would approach to within its tidal disruption periapsis of the \ac{MBH} within 1000 outer orbits.  A tail of longer time delays is omitted through this choice; see Section \ref{sec:discussion}.}
\label{fig:Time_TDMP}
\end{figure}

Among the simulated population of binaries most still require $\sim 100$ orbits after merger before the merger product is disrupted.   
Figure \ref{fig:Time_TDMP} shows the distribution of the time between the tidally stimulated merger and the tidal disruption of the merger product, under the assumption that the merger itself does not alter the trajectory around the \ac{MBH}.  Merger products can be disrupted in as little as $\sim 10^5$ years.  The inclusion of stellar tides does not appreciably change this distribution.

\begin{figure}
\centering
\includegraphics[width=\linewidth]{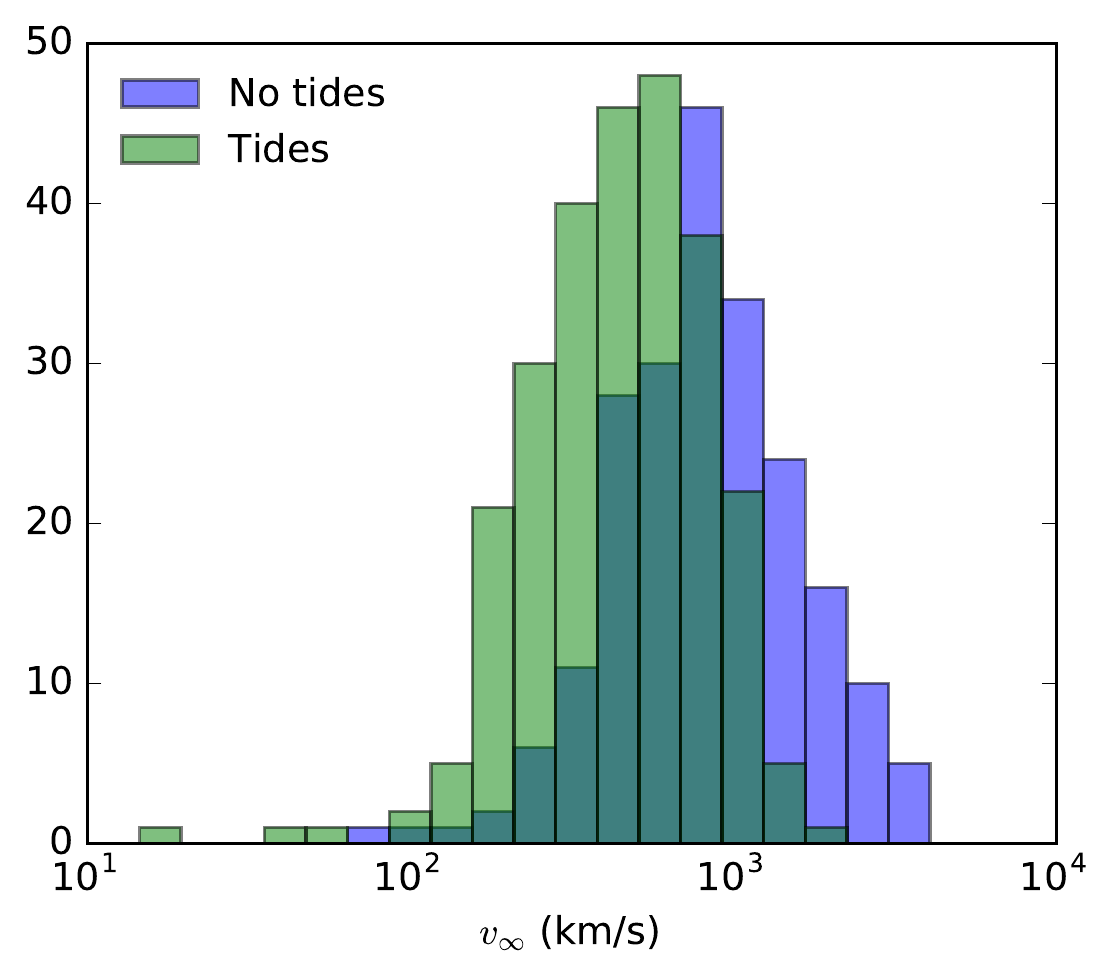}
\caption{The distribution of hyperbolic excess velocities of tidally ejected stars.}
\label{fig:hypervel}
\end{figure}

Although mergers are more common than binary tidal separations in our simulations, a considerable number of \ac{HVS} are still produced. The distribution of hyperbolic excess velocities of the ejected stars in tidally separated binaries $v_{\infty}^2=v_{*}^{2}-2GM_{\mathrm{MBH}}/r$, where $v_{*}$ is the star's current velocity and $r$ is the radial distance from the star to the MBH, is shown in Figure \ref{fig:hypervel}. A number of stars are ejected with velocities exceeding $1000\,\mathrm{km\: s^{-1}}$.   Stellar tides generally reduce ejection velocities. 

\section{Discussion}
\label{sec:discussion}

We simulated the dynamics of stellar binaries in the empty loss cone around an \ac{MBH}.  These systems preferentially merge over becoming tidally separated as the tide on the binary from the \ac{MBH} efficiently drives eccentricity evolution without significantly affecting the binary's energy (see Figure \ref{fig:sma_vs_ecc}).  We found that stellar binaries on highly eccentric orbits that gradually diffuse in angular momentum produce mergers in $\gtrsim 75\%$ of our simulations.  The remaining binaries become tidally separated and produce \ac{HVS} with hyperbolic excess velocities up to $1000\,\mathrm{km\:s^{-1}}$.   Consequently, more than ten percent of all tidal disruption events should be disruptions of merger products.  

Such mergers may generate strong magnetic fields through a dynamo mechanism \citep{ref:Wickramasinghe_et_al_2014}.  When the merger products are subsequently tidally disrupted by the \ac{MBH}, these strong magnetic fields could play a crucial role in powering the prompt formation of jets such as observed in Swift J164449.3+573451 \citep{GianniosMetzger:2011}.

Stellar tides between the binary components slightly reduce the merging fraction, and significantly reduce the eccentricity of the binary at the time of merger for many systems, thereby reducing the collision velocity.  We used an equilibrium tide model that only includes quadrupolar terms in the stellar deformation and loses accuracy at high eccentricity.  A more careful treatment of tides would include a dynamical tide model \citep[e.g.,][]{ref:Fabian_et_al_1975,ref:Zahn_1977,ref:Press_Teukolsky_1977} and additional effects from stellar rotation and tides raised by the \ac{MBH}.  This would increase tidal efficiency for very eccentric systems, possibly further reducing the eccentricity and relative velocity at merger.

The treatment of the collision could be improved with more detailed hydrodynamical simulations to more accurately determine the structure of the merger product \citep{ref:Antonini_et_al_2011}, the mass loss from the system, and any change in the trajectory of the merger product following merger, which we have ignored in our simplified analysis.  Moreover, Roche lobe overflow as the stars approach prior to merger could lead to a softer collision.  

\citet{ref:Stephan_et_al_2016} found that stellar evolution played an important role in the evolution of 65\% of the binaries they simulated in orbit around an \ac{MBH}. However, stellar evolution is unlikely to significantly affect our results, as these systems evolve on a typical timescale of $\sim\mathrm{Myrs}$.

The shortest-period binaries with the highest orbital velocity are expected to produce the fastest hypervelocity stars when the binary is tidally separated by an \ac{MBH}.  However, our simulations show that the closest binaries are preferentially merged rather than tidally separated in the empty loss cone.  This depletes the high-velocity tail of the hypervelocity star distribution, especially when stellar tides are included.  The absence of the high-velocity tail appears consistent with observations of Galactic hypervelocity stars \citep{Rossi:2014,Rossi:2017}.  However, the tidally stimulated merger fraction is significantly lower for binaries in the full loss cone \citep{ref:Mandel_Levin_2015}, so full loss cone binaries could still provide contributions to the high-velocity tail unless other effects suppress tidal separations of close binaries in that regime.

Our simulations were selected to focus on binaries whose angular momentum random walk around the \ac{MBH} will bring the merger product to a sufficiently small periapsis for tidal disruption within 1000 orbits.  The actual fraction of merger products disrupted within a few million years is $\lesssim 10\%$.  A long delay could allow the merger product to return to equilibrium, reducing the stellar radius, but an enhanced magnetic field could be retained in the radiative zones of the merger product \citep{BraithwaiteSpruit:2004}.  Moreover, many merger products will come sufficiently close to the \ac{MBH} for partial disruption long before full tidal disruption, and the extended merger products will be particularly susceptible to partial disruptions.  \citet{GuillochonMcCourt:2017} argue that such disruptions can increase the magnetic field by factors of $\sim 20$.  The combined effect of a merger followed by repeated partial disruptions can have a truly dramatic effect on the star's magnetic field.  \citet{Tchekhovskoy:2014} argued that strong coherent fields are required for prompt jet formation.  \citet{Parfrey:2015} advocate a less stringent requirement for the large-scale field and the ability of small-scale fields to produce jets when interacting with a spinning black hole.  Clearly, strong initial fields prior to disc formation can enhance the magnetic activity of the disc and jet formation.  Tidally stimulated stellar mergers could thus be an important ingredient in the production of jets during subsequent \acp{TDE}.

\section*{Acknowledgements}
\label{sec:acknowledgements}

Simulations in this paper made use of the IAS15 N-body integrator \citep{ref:Rein_Spiegel_2015} found in the REBOUND package \citep{ref:Rein_Liu_2012}, which can be downloaded at \url{http://github.com/hannorein/rebound}.  We are grateful to Will Farr, James Guillochon, Elena Maria Rossi and Alison Farmer for useful discussions.  IM acknowledges partial support by the STFC and by the National Science Foundation under Grant No.~NSF PHY11-25915.  YL acknowledges research support by the Australian Research Council Future Fellowship.

\bibliographystyle{mnras}
\bibliography{references/tidal_disruptions_references.bib,references/Mandel.bib}

\bsp	
\label{lastpage}
\end{document}